\documentclass{article}
\usepackage{spconf,amsmath,graphicx}
\usepackage{hyperref}
\usepackage{balance}
\usepackage{amsmath,amsfonts}
\usepackage{algorithmic}
\usepackage{algorithm}
\usepackage{array}
\usepackage[caption=false,font=normalsize,labelfont=sf,textfont=sf]{subfig}
\usepackage{textcomp}
\usepackage{stfloats}
\usepackage{url}
\usepackage{verbatim}
\usepackage{graphicx}
\usepackage{cite}
\usepackage{multirow}
\usepackage{multicol}
\usepackage{makecell}
\usepackage{diagbox}
\usepackage{color,soul}
\usepackage{ctable} 
\usepackage{flushend}

\title{A Conformer Based Acoustic Model For Robust Automatic Speech Recognition}
\name{Yufeng Yang$^1$, Peidong Wang$^{1*,2}$, and DeLiang Wang$^{1, 3}$\thanks{*This work was done when Peidong Wang was at The Ohio State University.}}
\address{$^1$Department of Computer Science and Engineering, The Ohio State University, USA\\
$^2$Microsoft Corporation, USA\\
$^3$Center for Cognitive and Brain Sciences, The Ohio State University, USA\\
{\href{mailto:yang.5662@osu.edu}{\texttt{yang.5662@osu.edu}}, \href{mailto:peidongwang@ieee.org}{\texttt{peidongwang@ieee.org}}, \href{mailto:wang.77@osu.edu}{\texttt{wang.77@osu.edu}}}
}
\begin{document}
\ninept
\maketitle
\begin{abstract}
This study addresses robust automatic speech recognition (ASR) by introducing a Conformer-based acoustic model. The proposed model builds on the wide residual bi-directional long short-term memory network (WRBN) with utterance-wise dropout and iterative speaker adaptation, but employs a Conformer encoder instead of the recurrent network. The Conformer encoder uses a convolution-augmented attention mechanism for acoustic modeling. The proposed system is evaluated on the monaural ASR task of the CHiME-4 corpus. Coupled with utterance-wise normalization and speaker adaptation, our model achieves $6.25\%$ word error rate, which outperforms WRBN by $8.4\%$ relatively. In addition, the proposed Conformer-based model is $18.3\%$ smaller in model size and reduces total training time by $79.6\%$.
\end{abstract}

\section{Introduction}
\label{sec:intro}
Riding on the tremendous success of deep learning, automatic speech recognition (ASR) technology has seen rapid advances in recent years, and is now widely deployed in voice-based human computer interaction. On the other hand, in far-field noisy environments, ASR performance degrade substantially \cite{haeb2020far}. Designing robust ASR systems in such conditions remains a technical challenge in real-world applications \cite{barker2015third, vincent2017analysis, barker2018fifth}.

Modern ASR is mostly based on deep neural networks (DNNs). ASR systems consist of acoustic modeling and language model based decoding. Hybrid systems usually use DNN to estimate the state of each time frame of an input signal such as senone states, and hidden Markov models to decode the state information into final transcripts \cite{trentin2001survey, hinton2012deep, dahl2011context, rabiner1986introduction}. Recently end-to-end (E2E) models have been used to directly estimate a word transcript without HMM-based decoding, usually using connectionist temporal classification (CTC) or recurrent neural network transducer (RNN-T)  \cite{graves2014towards, graves2006connectionist, zhang2020transformer, rao2017exploring}. Toolkits such as ESPnet provide platforms for E2E ASR, as well as strong benchmark models on various corpora \cite{watanabe2018espnet, watanabe20212020}. However, on the widely used CHiME-4 corpus \cite{vincent2017analysis}, E2E methods in ESPnet do not outperform Kaldi-based hybrid systems \cite{povey2011kaldi}. In this study, we focus on developing a robust acoustic model for the monaural ASR task of the CHiME-4 corpus, with modules that are primarily used in E2E ASR models.

The wide residual BLSTM network (WRBN) achieved the top-rank performance on the monaural speech recognition task in the original evaluation of the CHiME-4 challenge \cite{jahn2016wide}. Subsequently, adding utterance-wise dropout and iterative speaker adaptation leads to a considerable improvement over the original WRBN in terms of word error rate (WER) \cite{wang2018utterance,wang_filter-and-convolve_2018}. The WER performance was further improved by employing an LSTM language model (LSTMLM) \cite{wang2020complex}. However, this system is inefficient because of the recurrent nature of BLSTM, requiring long model training time.

The Transformer model \cite{vaswani2017attention} uses an attention mechanism to represent temporal contexts of an entire sequence, and it has been shown to outperform recurrent neural networks for ASR tasks \cite{karita2019comparative, zhang2020transformer}. Although Transformer is capable of leveraging temporal information over long sequences, overcoming a drawback of BLSTM, its ability to leverage local information in a sequence appears limited. To deal with this issue, a convolution-augmented Transformer model, named Conformer \cite{conformer}, was proposed, and it produces better results than Transformer-based models on ASR and other tasks \cite{guo2021recent, chen2021continuous}. 

In this paper, we propose to integrate the WRBN and Conformer encoder into a new Conformer-based acoustic model for monaural robust ASR. We apply utterance-wise processing to all normalization layers in our model, which leads to more reliable computation for each utterance and avoids inter-utterance interference. The utterance-wise Layernorm (LN) used in this study can also be applied to all LN layers that process batches with padding, when the knowledge of feature length is available \emph{a priori}. We also utilize iterative speaker adaptation for post-processing. Evaluated on the CHiME-4 corpus, the proposed model outperforms WRBN by $8.4\%$ relatively in terms of WER. In addition, the size of our model is reduced by $18.3\%$, and total training time is cut by $79.6\%$.

The remainder of the paper is organized as follows. Details of the proposed system are described in Section~\ref{sec:am}. Experimental setup and evaluation results are presented in Section~\ref{sec:exp} and Section~\ref{sec:results}, respectively. Section~\ref{sec:conclusions} concludes the paper.

\section{System Description}
\label{sec:am}

\subsection{Wide Residual BLSTM Network}
In this study, wide residual convolutional layers in WRBN before BLSTM layers are utilized and denoted as WRCNN. WRCNN passes the input signal through a convolution layer and uses three residual blocks to extract representation at different frequency resolutions \cite{zagoruyko_wide_2016}. Afterwards an utterance-wise Batchnorm (BN) and a linear layer with ELU (exponential linear unit) non-linearity are utilized to project the signal into proper dimensions.

\subsection{Conformer Encoder}
A Conformer encoder builds upon a Transformer encoder, with an additional convolutional network, and macaron-like feed-forward layers \cite{conformer}. We employ a Conformer encoder to leverage sequence information via convolution-augmented attention. Modules inside the Conformer encoder are described below.

\subsubsection{Feed-forward Network}\label{sec:ffn}
Feed-forward network (FFN) contains two linear layers with an activation function between them, and residual connections over the entire module. We adopt the pre-norm architecture in \cite{xiong2020layer}. So the FFN is defined as

\begin{equation}
\begin{aligned}
\text{FFN}(\mathbf{x}) = \mathbf{W}_{2}\text{Dropout}(\text{Swish}(\mathbf{W}_{1}\mathbf{x} + \mathbf{b}_{1})) + \mathbf{b}_{2},
\end{aligned}
\end{equation}
where $\mathbf{x}$ is the input signal, Swish is an activation function defined as $\text{Swish}(x) = x\text{sigmoid}(x)$. $\mathbf{W}_{1}\in\mathbb{R}^{d_{\text{ff}}\times d_{\text{attn}}}$, $\mathbf{b}_{1}\in\mathbb{R}^{d_{\text{ff}}}$, and $\mathbf{W}_{2}\in\mathbb{R}^{d_{\text{attn}}\times d_{\text{ff}}}$, $\mathbf{b}_{2}\in\mathbb{R}^{d_{\text{attn}}}$ are weights and biases for the first and second linear layers, respectively. The first linear layer expands attention dimension by a factor of $4$, i.e. $d_{\text{ff}} = 4d_{\text{attn}}$, where $d_{\text{attn}}$ is the attention dimension.

By applying utterance-wise LN on the input as pre-norm, each FFN module processes the signal as

\begin{equation}
\begin{aligned}
\text{Output} = \mathbf{x} + \frac{1}{2}\text{Dropout}(\text{FFN}(\text{LN}({\mathbf{x}}))).
\end{aligned}
\end{equation}

\subsubsection{Multi-head Self-attention Network}


In the multi-head self-attention (MHSA) network, absolute positional encoding is employed to encode positional information in the sequence. In this study, the same positional encoding as in \cite{vaswani2017attention} is applied without scaling the input signal. Instead we divide the positional encoding matrix by the scaling factor. That is, before the MHSA module, instead of 
\begin{equation}
    \begin{aligned}
    \text{Output} = \sqrt{d_{\text{attn}}} \mathbf{x} + \mathbf{PE},
    \end{aligned}
\end{equation}
we use

\begin{equation}
    \begin{aligned}
    \text{Output} = \mathbf{x} + \frac{1}{\sqrt{d_{\text{attn}}}}\mathbf{PE},
    \end{aligned}
\end{equation}
where $\mathbf{x}$ is the input signal, $\mathbf{PE}$ is the positional encoding matrix, and $\sqrt{d_{\text{attn}}}$ is the scaling factor. Our modification here essentially avoids the scaling of the original input feature.

For each head $h$ in the MHSA, self-attention is computed as
\begin{equation}
\begin{aligned}
\text{Attention}(\mathbf{Q}_{h}, \mathbf{K}_{h}, \mathbf{V}_{h}) = \text{Softmax}(\frac{\mathbf{Q}_{h}\mathbf{K}_{h}^{T}}{\sqrt{d_{\text{attn}}}})\mathbf{V}_{h},
\end{aligned}
\end{equation}
where $\mathbf{Q}_{h}=\mathbf{W}_{Q}^{h}\mathbf{x}$, $\mathbf{K}_{h}=\mathbf{W}_{K}^{h}\mathbf{x}$, and $\mathbf{V}_{h}=\mathbf{W}_{V}^{h}\mathbf{x}$ are query, key, and value linear projections of head $h$ on the input sequence, respectively. $\mathbf{W}_{Q}^{h}$, $\mathbf{W}_{K}^{h}$, and $ \mathbf{W}_{V}^{h}\in\mathbb{R}^{ \frac{d_{\text{attn}}}{H}\times d_{\text{attn}}}$ are projection weights for query, key, and value, respectively. $H$ is the number of heads.

After the computation of self-attention, all the heads are concatenated and fed to a final linear layer,

\begin{equation}
\begin{aligned}
\text{MHSA}(\mathbf{Q}, \mathbf{K}, \mathbf{V}) & = \textbf{W}_{\text{out}}\text{Concat}(\text{head}_{1},...,\text{head}_{H}),
\end{aligned}
\end{equation}
where $\mathbf{W}_{\text{out}}\in\mathbb{R}^{d_{\text{attn}}\times d_{\text{attn}}}$ is the weight matrix of the final linear layer, and $\text{head}_{h} = \text{Attention}(\mathbf{Q}_{h}, \mathbf{K}_{h}, \mathbf{V}_{h})$.

Therefore, each MHSA module processes the signal as:

\begin{equation}
\begin{aligned}
\text{Output} = \mathbf{x} + \text{Dropout}(\text{MHSA}(\text{LN}({\mathbf{x}}))).
\end{aligned}
\end{equation}

\subsubsection{Convolutional Network}

With similar architecture as in \cite{conformer}, the convolutional network consists of a pointwise convolution, followed by a GLU (gated linear unit) activation function. After a 1-dimensional depthwise convolution and an utterance-wise BN, the Swish activation is applied. Finally a 1-dimensional pointwise convolution is employed. Note that all convolutions operate on the time dimension. 

\subsection{Utterance-wise Normalization}
Inspired by the utterance-wise dropout in \cite{wang2018utterance}, we modify all normalization layers into utterance-wise normalization. BN in the convolutional network is substituted by the same utterance-wise BN as in WRBN \cite{jahn2016wide}. Statistics over the time dimension are collected to normalize the feature dimension for each utterance, treating batch as an independent dimension. Utterance-wise BN can obtain more reliable estimation of statistics from each utterance without including other utterances, and also makes the test stage more independent of the batch constellation in the training stage.

In this study, LN in each module of the Conformer encoder is modified into utterance-wise LN, which is defined as

\begin{equation}
\begin{aligned}
\mathbf{y} = \frac{\mathbf{x} - \mu}{\sqrt{\sigma^2 + \epsilon}}\cdot \boldsymbol{\gamma} + \boldsymbol\beta,
\end{aligned}
\end{equation}
where $\mathbf{x}$ and $\mathbf{y}$ are the input and output of the LN layer. $\mu$ and $\sigma$ are the mean and standard deviation of the hidden layer units. $\boldsymbol{\gamma}$ and $\boldsymbol\beta$ are learnable affine transform parameters, and $\epsilon$ is a small number for numerical stability.

\begin{figure}[h!]
    \centering
    \includegraphics[width=0.7\linewidth]{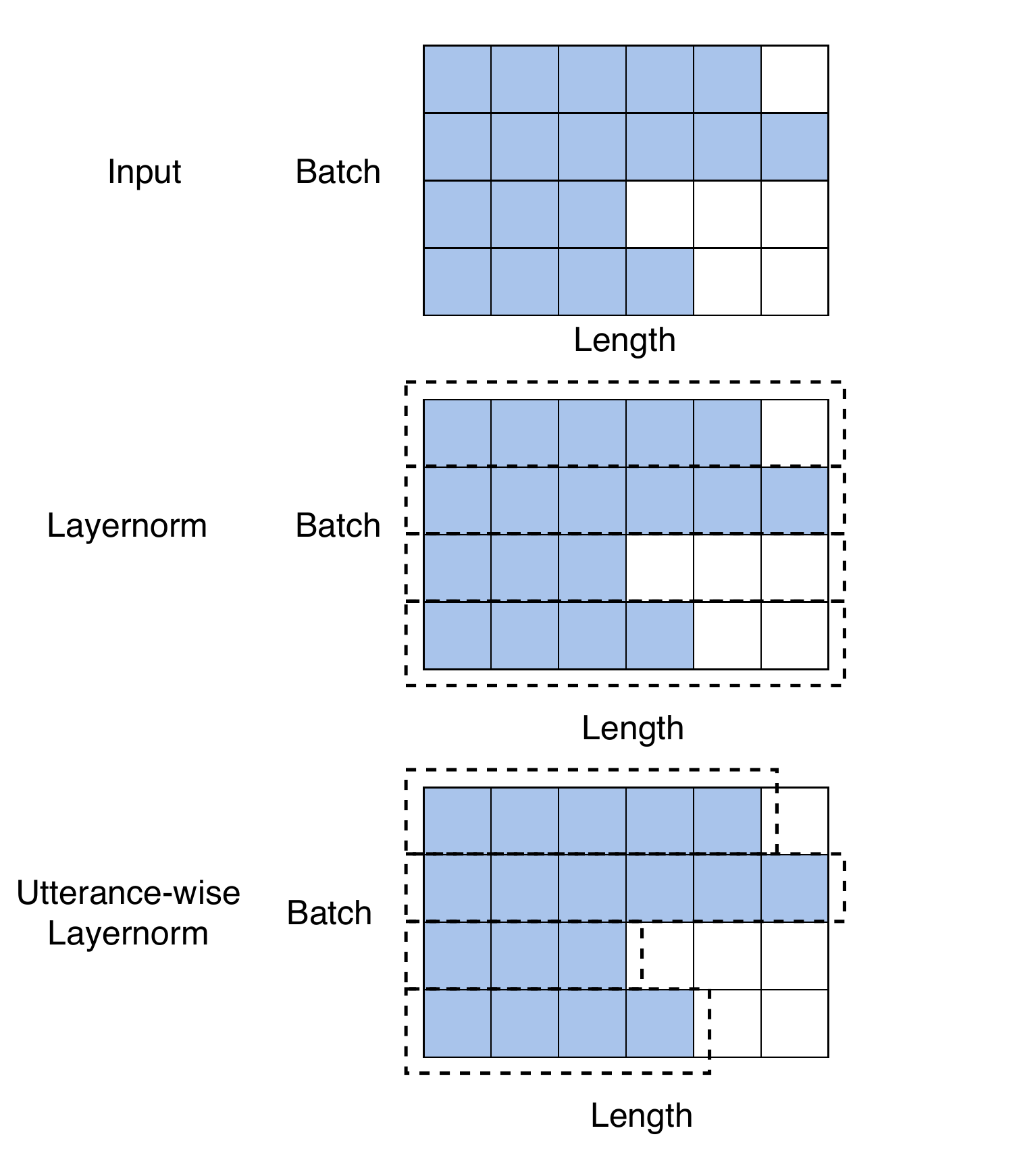}
    \caption{Illustration of original Layernorm and proposed utterance-wise Layernorm.}
    \label{fig:layernorm}
\end{figure}

Fig.~\ref{fig:layernorm} shows how utterance-wise LN work. For illustration purposes, we assume the feature dimension to be 1 and maximum utterance length of batch to be 6. Given an input batch size 4, blue cells indicate the feature for each utterance and white cells indicate the zero padding. Each utterance starts from left to right. Dashed lines surrounding each utterance include the features subject to LN processing. For LN, during the normalization of a hidden layer, i.e. the feature dimension, zero padding is included when updating the parameters $\boldsymbol\gamma$ and $\boldsymbol\beta$. In this case, the entire normalized feature output will be biased due to the zero padding, degrading the reliability of the collected statistics. The proposed utterance-wise LN overcomes this bias by processing features taking into account of the utterance length. For each utterance, only the useful features will be included for updating parameters. As a result, zero padding will not affect the estimation of learnable parameters in utterance-wise processing, resulting in more precise computation and normalization. 

In addition to utterance-wise normalization, we apply sequence masking in every step of the computation. For example, after each linear layer in FFN, we apply an all-zero mask over zero padding to remove the bias introduced during computation. Also, for an utterance with zero padding, when computing attention weights, we mask out the padding parts such that attention is computed only between valid time frames. Therefore, we believe that utterance-wise normalization and masking benefit training stability and convergence.

\subsection{Model Architecture}
\label{sec:model_arch}
\begin{figure}[htbp]
    \centering
    \includegraphics[width=0.95\linewidth]{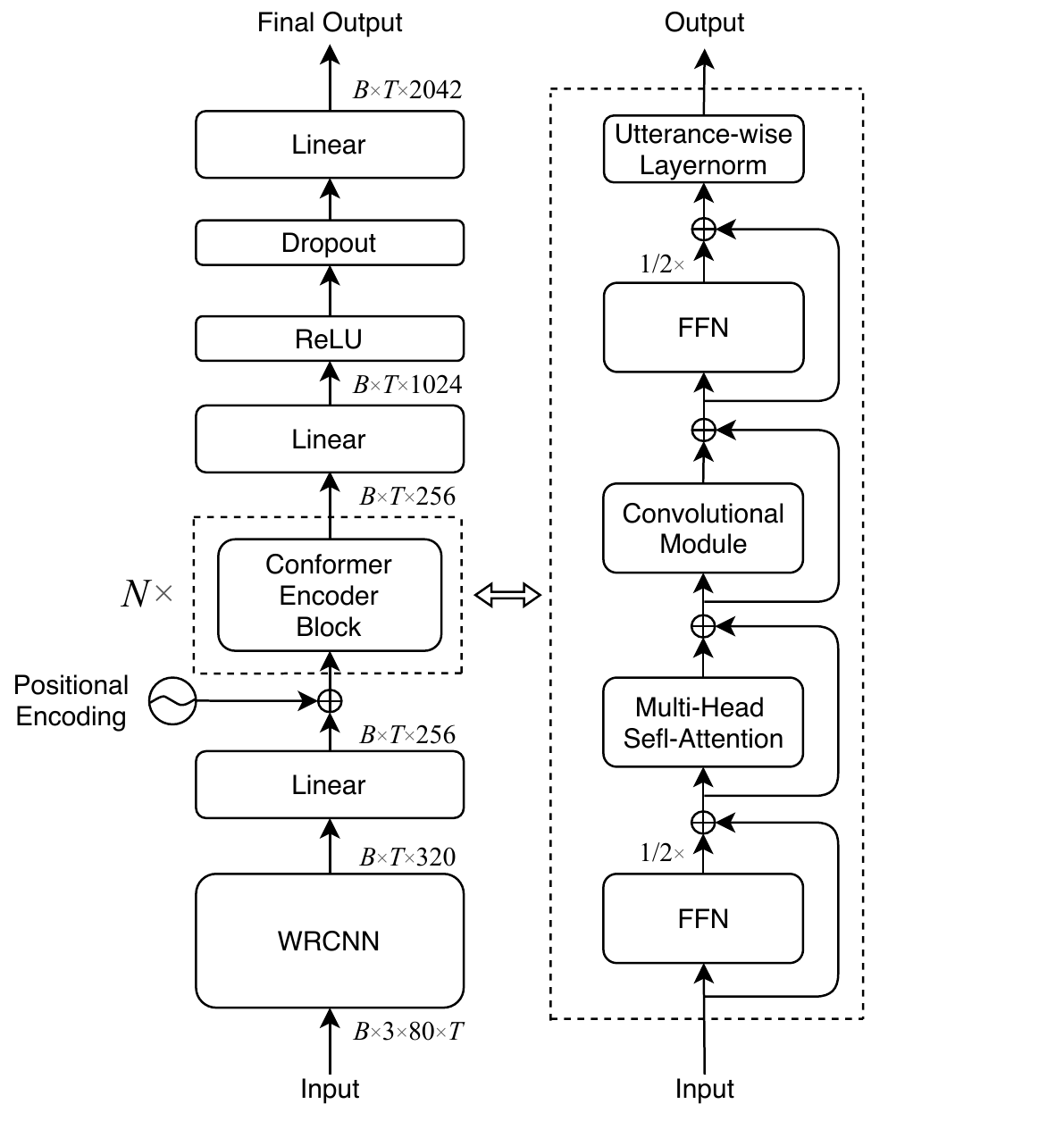}
    \caption{Model architecture of Conformer-based acoustic model. $B$ denotes batch size, and $T$ denotes number of time frames of the longest utterance in a batch.}
    \label{fig:model}
\end{figure}

Our proposed acoustic model integrates the WRCNN and Conformer encoder, and uses utterance-wise normalizations. Fig.~\ref{fig:model} shows the acoustic model architecture. The input signal is $80$-dimensional mean-normalized log-Mel filterbank features extracted from the original single-channel noisy speech signal, coupled with its delta and delta-delta features. It is processed by WRCNN and projected to the dimension of MHSA by a linear layer. After $N$ blocks of the Conformer encoder with absolute positional encoding, the signal is projected into $1024$ dimensions, followed by a ReLU (rectified linear unit) activation and dropout. Finally a linear layer projects the signal to the final output of each frame as posterior probability for $2042$ context-dependent states.

\section{Experimental Setup}
\label{sec:exp}
\subsection{CHiME-4 Corpus}
The CHiME-4 corpus \cite{vincent2017analysis} has six-microphone real and simulated noisy speech recordings. Real recordings are made in noisy environments (bus, cafe, pedestrian area, and street junction). Simulated data are noisy utterances generated by artificially mixing clean speech data with noisy backgrounds. The main goal is to recognize the real speech data in CHiME-4 challenge.

The training data has $1600$ real and $7138$ simulated recordings for each channel, from a total of $4$ speakers in the real data, and 83 speakers from the WSJ$0$ SI-$84$ training set in the simulated data. The development data consists of 1640 real and 1640 simulated utterances from $4$ other speakers not included in the training data. The test data has $1320$ real and $1320$ simulated utterances from another $4$ different speakers. In this study, single-channel speech signals from six individual microphones are all utilized to train the acoustic model. The development and test utterances are randomly selected from six microphones. The monaural ASR task uses one of the six microphones for testing.

\subsection{Implementation Details}
The configuration of WRCNN is kept the same as described in \cite{jahn2016wide}. The layer number $N$ of the Conformer encoder is set to $2$, and attention dimension $d_{\text{attn}}$ is set to $256$. The kernel size of the $1$-D depthwise convolution is set to $16$. We use the same Transformer learning rate schedule as in \cite{vaswani2017attention}, with $20$k warm-up steps and the learning rate factor of $5$. The Adam optimizer with $\beta_{1}=0.9$, $\beta_{2}=0.98$, and $\epsilon=1e^{-9}$ is utilized for model training and fine-tuning. The batch size is $4$ and short utterances are padded with zeros to match the length of the longest utterance in each batch. Based on the model performance on development data, we fine-tune the best performing model (usually found within 10 epochs) using exponential moving average with the decay factor set to $0.999$ and initial learning rate of the optimizer is set to $1e^{-5}$. The dropout rate is set to $0.15$ for network and attention weights. In our replication of WRBN, the learning rates for training and fine-tuning are set to $1e^{-4}$ and $1e^{-5}$, respectively. The 2042 senone labels for the training data are generated from the provided GMM-HMM system in the original CHiME-4 challenge.

For decoding implementation, we follow the same steps as described in \cite{wang2020complex}. The decoding beamwidth is set to be $180$, lattice beamwidth is $120$, number of active tokens ranges from $20000$ to $80000$. For iterative speaker adaptation, a linear input network (LIN) is utilized. The Adam optimizer is used with the initial learning rate of $10^{-4}$. We train the linear layer for $10$ epochs for each speaker in the simulated and real data. The linear layer for LIN is initialized into an $80\times80$ identity matrix, and the rescored result from the language model is used for the next iteration. After the first iteration, the linear layer for LIN will be re-initialized and trained again. In our implementation, iterative adaptation is applied $3$ times. In the computation of WER, language model weights ranging from 4 to 25 are selected according to the performance on the development data for language model rescoring and speaker adaptation.

\section{Evaluation and Comparison Results}
\label{sec:results}

\subsection{Results and Comparisons}
Our proposed Conformer-based acoustic model achieves $6.25\%$ WER on the CHiME-4 real test data. To our knowledge, this WER result represents the best score on the monaural CHiME-4 task following the official CHiME-4 recipe. Table~\ref{tab:asr_comp} lists the results of our system along with several other baseline models. The baseline models include the Kaldi hybrid system based on a time delay neural network (TDNN) with lattice-free maximum mutual information and LSTMLM \cite{chen2018building}, ESPnet E2E Conformer based model \cite{guo2021recent}, original WRBN \cite{jahn2016wide}, Du \emph{et al.}'s original top-rank system on the CHiME-4 challenge that uses deep convolution neural networks and LSTMLM by \cite{du2016ustc}, Chen \emph{et al.}'s ASR system using triplet-loss based embedding and a factored form of TDNN \cite{chen_2021_scenario}, and Wang \emph{et al.}'s previous best system using WRBN with utterance-wise dropout and iterative speaker adaptation \cite{wang2020complex}. The row of \emph{Mixtures + Tri-gram} denotes the performance achieved using the standard tri-gram language model (LM) with noisy speech as input. After five-gram LM and RNN LM (RNNLM) rescoring, the WERs are improved as shown in the next row (\emph{Five-gram and RNNLM}). The results get further improved by replacing RNNLM with LSTMLM in the next row (\emph{Five-gram and LSTMLM}). After three iterative speaker adaptations, the row of \emph{Iterative Speaker Adaptation} provides the final results of our ASR system.

\begin{table}[htbp!]
    \centering
    \caption{ASR performance (\%WER) of the proposed model and other comparison systems.}
    \label{tab:asr_comp}
    \centering
    \scalebox{0.85}{
    \begin{tabular}[width=\linewidth]{ l | c | c | c | c }
         \hline
         
          \makecell[c]{\multirow{2}{*}{System}} & \multicolumn{2}{c|}{Dev. Set} &  \multicolumn{2}{c}{Test Set} \\
         \cline{2-5}
          & Simu. & Real & Simu. & Real \\
         \hline
         Mixtures + Tri-gram & 9.33 & 7.08 & 13.85 & 11.72\\
         \hline
         \hspace{1mm} + Five-gram and RNNLM & 7.31 & 5.11 & 11.32 & 9.10 \\
         \hline
         \hspace{1mm} + Five-gram and LSTMLM & 5.99 & 4.16 & 10.63 & 8.15\\
          \hline
         \hspace{2mm} + Iterative Speaker Adaptation & \textbf{4.99} & \textbf{3.35} & \textbf{8.61} & \textbf{6.25} \\
         \specialrule{.15em}{0em}{0em}
          Kaldi Baseline \cite{chen2018building} & 6.81 & 5.58 & 12.15  & 11.42 \\
           \hline
           E2E Conformer \cite{guo2021recent} & 9.10 & 7.90 & 14.20 & 13.40 \\
         \hline
         Original WRBN \cite{jahn2016wide} & 6.69 & 5.19 & 11.11 & 9.34\\
         \hline
         Du \emph{et al.} \cite{du2016ustc} & 6.61 & 4.55 & 11.81 & 9.15 \\
          \hline
         Chen \emph{et al.} \cite{chen_2021_scenario} & 6.55 & 4.43 & 12.03 & 10.29\\
         \hline
         Wang \emph{et al.} \cite{wang2020complex} & \textbf{4.99} & 3.54 & 9.41 & 6.82 \\
          \hline
         
    \end{tabular}}
\end{table}

As shown in Table~\ref{tab:asr_comp}, our proposed model outperforms the previous best system \cite{wang2020complex} by $8.4\%$ and $8.5\%$ relatively in real and simulated data, respectively. Meanwhile, our proposed model outperforms other systems by considerable margins. It is worth noting that ESPnet Conformer based E2E baseline cannot even outperform the Kaldi-based hybrid model baseline. The comparison listed in the table demonstrate that our system produces better WER results than both hybrid and E2E systems. We should note that recent studies based on self-supervised learning \cite{chang2022end, wang2022wav2vec, wang2022improving, zhu2022joint} have achieved better performance than the proposed system, but these systems leverage out-of-domain data and hence do not conform with the CHiME-4 evaluation recipe.


\subsection{Results in Different Noisy Environments}
Table~\ref{tab:asr_env} compares the proposed model with the previous best system \cite{wang2020complex} in all four noisy environments: Bus, Cafe, Pedestrian, and Street. The results illustrate that our proposed model is more robust for both simulated and real data in the Bus, Cafe, and Pedestrian environments; for the Street environment, Wang \emph{et al.}'s system produces slightly better WER.

\begin{table}[htbp!]
    \centering
    \caption{Comparisons of ASR performance (\%WER) in different noisy environments.}
    \label{tab:asr_env}
    \centering
    \begin{tabular}[width=\linewidth]{ c | c | c | c | c }
         
         \hline
          \makecell[c]{\multirow{2}{*}{Environments}} & \multicolumn{2}{c|}{Wang \emph{et al.} \cite{wang2020complex}} &  \multicolumn{2}{c}{Proposed} \\
         \cline{2-5}
           & Simu. & Real & Simu. & Real \\
         \hline
         Bus & 6.76 & 10.66 & \textbf{5.90} & \textbf{9.01}\\
         \hline
         Cafe & 11.34 & 7.08 & \textbf{9.97} & \textbf{6.44} \\
         \hline
         Pedestrian & 8.42 & 5.10 & \textbf{7.38} & \textbf{5.03}\\
          \hline
         Street & \textbf{11.13} & \textbf{4.46} & 11.19 & 4.52 \\
         \specialrule{.15em}{0em}{0em}
        Average & 9.41 & 6.82 & \textbf{8.61} & \textbf{6.25}\\
         \hline
         
    \end{tabular}
\end{table}

\subsection{Training Efficiency and Model Size}
Table~\ref{tab:comp_size} compares the proposed system with the previous best system \cite{wang2020complex} in terms of training efficiency and model size. Both models are trained on one NVIDIA Volta V100 GPU with the same batch size. The WRBN model is trained for 6 epochs and fine-tuned for 4 epochs, while the proposed acoustic model is trained for 5 epochs and fine-tuned for 11 epochs. Percentage of reduction is shown in the last row for each measure. From Table~\ref{tab:comp_size} we can see that, our proposed model achieves better training efficiency by cutting total training time by $79.6\%$, and average training time by $87.3\%$, and reduces model size by $18.3\%$.

\begin{table}[htbp!]
    \centering
    \caption{Comparisons of training efficiency and model size.}
    \label{tab:comp_size}
    \centering
    \scalebox{0.95}{
    \begin{tabular}[width=\linewidth]{ c | c | c | c }
         \hline
         
          Approaches &  \makecell[c]{Total\\ Training \\Time\\(hour)} & \makecell[c]{Avg. \\Training\\ Time\\ (hour/epoch)} & \makecell[c]{Model Size\\ (MB)}\\
         \hline
        Wang \emph{et al.} \cite{wang2020complex} & 105.20& 10.52 &66.63\\
         \hline
        Proposed  & 21.48 & 1.34 & 54.44 \\
        \specialrule{.15em}{0em}{0em}
         Reduction (\%)& 79.6 & 87.3 & 18.3\\
         \hline
         
    \end{tabular}}
\end{table}

\section{Concluding Remarks}
\label{sec:conclusions}
In this paper, a Conformer-based acoustic model is proposed for robust ASR, and evaluated on the monaural task of the CHiME-4 challenge. Coupled with utterance-wise normalization and iterative speaker adaptation, our proposed model achieves $6.25\%$ WER on the real test data, outperforming the previous best system by $8.4\%$ relatively. In addition, compared to the baseline system, the proposed model is $18.3\%$ smaller in model size and takes $79.6\%$ less training time. 

In the proposed system, the number of Conformer encoder blocks is set to $2$, while other studies usually use more than $6$ blocks. Our experiments with more encoder blocks do not produce better results. It is likely that the data in the CHiME-4 corpus is not sufficient to train models with more encoder blocks. Thus the performance of the proposed system may be further improved by augmenting training data, which represents one of future research directions. Future work also includes adding a speech enhancement frontend for monaural ASR and extending the acoustic model to multi-channel cases.

\section{Acknowledgements}
\label{sec:acknowledgement}
This research was supported in part by a National Science Foundation grant (ECCS-1808932) and the Ohio Supercomputer Center.

\bibliographystyle{IEEEtran}
\bibliography{strings,refs}

\end{document}